\documentclass[
onecolumn,
preprint,
showpacs,
preprintnumbers,
nofootinbib,
 amsmath,amssymb,
 prd,
]{revtex4-1}
\usepackage{cases}
\usepackage{graphicx}% Include figure files
\usepackage{dcolumn}% Align table columns on decimal point
\usepackage{bm}
\usepackage{dcolumn,color}% bold math
\begin{document}

\title{Linearization of a warped $f(R)$ theory in the higher-order frame}

\author{Yuan Zhong$^{1}$, Yu-Xiao Liu$^{2}$\footnote{Corresponding author: liuyx@lzu.edu.cn}}
\affiliation{$^1$
    School of Science, Xi'an Jiaotong University, Xi'an 710049, People's Republic of China}
\affiliation{$^2$Institute of Theoretical Physics, Lanzhou University,
           Lanzhou 730000, People's Republic of China}

\begin{abstract}
The linearization of a type of $f(R)$ gravity is studied directly in the higher-order frame for an arbitrary five-dimensional warped space-time background. The quadratic actions of the normal modes of the scalar, vector, and tensor perturbations are derived by taking the curvature gauge, under which the linear perturbation of the scalar curvature is zero, and all the perturbation equations reduce to second order. By comparing our results to those obtained in the Einstein frame, we find that the quadratic actions of the normal modes are equivalent for these two frames.
\end{abstract}

% PACS, the Physics and Astronomy
                             % Classification Scheme.
%\keywords{Suggested keywords}%Use showkeys class option if keyword
                              %display desired
\maketitle

%\tableofcontents
\section{Introduction}
$f(R)$ gravity is one of the simplest extensions of Einstein's general relativity (GR). Its Lagrangian is a function of the scalar curvature $R$, and GR is simply the case with $f(R)\propto R$. The Einstein equations of an $f(R)$ theory are fourth-order differential equations, which are hard to solve in general. Nevertheless, $f(R)$ theory is mathematically related to the Brans-Dicke theory with parameter $\omega_0=0$~\cite{SotiriouFaraoni2010} (also known as ``massive dilaton gravity"~\cite{OHanlon1972,Wands1994}), which leads to second-order equations. In addition, a Brans-Dicke theory can always be transformed into a minimally coupled scalar-Einstein theory~\cite{SotiriouFaraoni2010,Wands1994,FaraoniGunzigNardone1999,Maeda1989,CapozzielloRitisMarino1997}, as can the original $f(R)$ theory~\cite{BarrowCotsakis1988}. In fact, in addition to the $\omega_0=0$ Brans-Dicke theory and the minimally coupled scalar-Einstein theory, an $f(R)$ theory has infinite representations which are mutually related via conformal transformations~\cite{Flanagan2004,SotiriouFaraoniLiberati2008}.
Following Ref.~\cite{SotiriouFaraoni2010}, we call the Brans-Dicke theory (the scalar-Einstein theory) as the Jordan frame (the Einstein frame) of the corresponding $f(R)$ theory, while we refer to the original $f(R)$ theory as the higher-order frame.\footnote{Sometimes, however, the original $f(R)$ theory is also called the Jordan frame.}

Physically, $f(R)$ theory offers a pure geometric explanation for cosmological inflation and dark energy~\cite{Starobinsky1980,BarrowOttewill1983,NojiriOdintsov2003d,CarrollDuvvuriTroddenTurner2004,CapozzielloCarloniTroisi2003,NojiriOdintsov2006} [see~\cite{DeTsujikawa2010} for discussions of $f(R)$ theory and its cosmological phenomenology] as well as the formation of domain wall branes in higher-dimensional space\footnote{In general relativity, one usually constructs a domain wall brane model by introducing a background scalar field (for example see~\cite{Gremm2000,DeWolfeFreedmanGubserKarch2000}). Since a pure metric $f(R)$ gravity is related to a minimally coupled scalar-Einstein theory, it is also possible to construct domain wall brane models by simply starting with a vacuum $f(R)$ theory.} ~\cite{ZhongLiu2016}.  Usually, the background solutions are found in the Jordan frame or in the Einstein frame, but sometimes it is not difficult to find interesting solutions directly in the higher-order frame (for example, see Ref.~\cite{ZhongLiu2016}). Once a solution is obtained in the Jordan or the Einstein frame, one can reconstruct the solution in the higher-order frame.

The mathematical equivalence between different frames was widely accepted in the past; only the physical equivalence remains a controversial issue (see~\cite{Sokolowski1989,Cotsakis1993,Schmidt1995,Teyssandier1995,Cotsakis1995,MagnanoSokolowski1994} for some of the early discussions and~\cite{CapozzielloRitisMarino1997,FaraoniGunzigNardone1999} for comprehensive reviews). However, it was recently pointed out that different frames are inequivalent even mathematically~\cite{Sk.Sanyal2016}. According to Ref.~\cite{Sk.Sanyal2016}, neither Noether equations nor quantum equations may be translated from one frame to the other because of the nontranslatability of momenta of different frames. Therefore, it is important to study $f(R)$ theory directly in its higher-order frame.

In addition to the construction of various kinds of background solutions, the evolution of gravitational perturbations around these solutions is also an important issue. On one hand, the perturbation equations should also be of fourth-order in the higher-order frame. While on the other hand, the conformal transformation technique indicates that the higher-order derivatives can be eliminated. To understand this ``paradox," one must confront the higher-order frame and derive the perturbation equations directly. This issue has been extensively considered in cosmology even for the more generalized $f(R,\phi)$ theory~\cite{Hwang1997a,Hwang1997}. In this paper, we consider the linear perturbations around a five-dimensional warped geometry in the higher-order frame of a pure metric $f(R)$ gravity.

Models with a warped extra dimension have been applied to explain the large hierarchy between the electroweak scale and the Planck scale~\cite{RandallSundrum1999,CabrerGersdorffQuiros2010,RaychaudhuriSridhar2016}, the splitting of fermion masses~\cite{GherghettaPomarol2000}, and the localization of four-dimensional gravity on a four-dimensional domain wall~\cite{RandallSundrum1999a,Gremm2000,DeWolfeFreedmanGubserKarch2000,CsakiErlichHollowoodShirman2000} (see~\cite{Quiros2015,Ponton2012} for recent reviews on the theory and phenomenology of warped spaces). It is a natural idea to consider warped spaces in more general $f(R)$ gravity~\cite{ZhongLiu2016,ParryPichlerDeeg2005,AfonsoBazeiaMenezesPetrov2007,HoffDias2011,LiuZhongZhaoLi2011,BazeiaMenezesPetrovSilva2013,BazeiaLobaoMenezesPetrovSilva2014,XuZhongYuLiu2015,YuZhongGuLiu2015}. In Ref.~\cite{ZhongLiuYang2011}, the tensor perturbation around an arbitrary $f(R)$ warped geometry was studied in the higher-order frame. Then in Ref.~\cite{ZhongLiu2016} all the perturbation modes (including scalar, tensor and vector modes) were investigated in the Einstein frame. The aim of the present work is to give a systematic analysis to the linearization of an arbitrary $f(R)$ warped geometry in the higher-order frame.

In the next section, we give a general consideration to the perturbation theory of a pure metric $f(R)$ gravity. The quadratic action of the metric perturbation will be derived for an arbitrary metric background. Five-dimensional $f(R)$ warped spaces will be considered in Sec.~\ref{secWarpedSpace}. Using the scalar-tensor-vector (STV) decomposition, we decompose the quadratic action into scalar, tensor and vector parts. Then in Sec.~\ref{secCurv} we derive the scalar perturbation equation in the curvature gauge and compare this result with the one obtained in the Einstein frame. A short summary will be given in Sec.~\ref{secSum}.
\section{Gravitational perturbations for arbitrary $f(R)$ background geometry}
\label{section2}
In this section, we consider the gravitational perturbations of a pure metric $f(R)$ gravity for an arbitrary $d$-dimensional background geometry.
We start with the action of a $d$-dimensional $f(R)$ gravity
\begin{eqnarray}
 \label{action}
  S=2M^{d-2}_\ast\int d^d x \sqrt{-\tilde{g}}f(\tilde{R}),
\end{eqnarray}
where $M_\ast$ is the fundamental Planck scale, $\tilde{g}=\det \tilde{g}_{MN}$ is the determinant of the metric $\tilde{g}_{MN}$ with indices $M,N=0,1,\cdots,d-1$, and $f(\tilde{R})$ is an arbitrary function of the Ricci scalar $\tilde{R}$.

Suppose the metric $\tilde{g}_{MN}$ can be split into a background $g_{MN}$ plus a small perturbation $H_{MN}$, such that
\begin{eqnarray}
\tilde{g}_{MN}&=&g_{MN}+H_{MN},\quad H_{MN}\ll g_{MN}.
\end{eqnarray}
Similarly, we can write the inverse of the metric as
\begin{eqnarray}
\tilde{g}^{MN}&=&g^{MN}+\delta g^{MN},
\end{eqnarray}
The demanding of  the orthogonal condition
\begin{eqnarray}
\tilde{g}^{MP}\tilde{g}_{PN}&=&\delta^M_{~N}=g^{MP}g_{PN},
\end{eqnarray}
leads to a relation between $\delta g^{MN}$ and $H_{MN}$
\begin{equation}
\label{EqFlucInversG}
H^{M}_N +{g}_{NP}\delta g^{MN}+H_{NP} \delta g^{MP}=0,
\end{equation}
where the indices of $H_{MN}$ are raised by the background metric $g^{MN}$:
\begin{equation}
H^{MN}\equiv g^{MP}g^{NQ}H_{PQ}.
\end{equation}
Equation \eqref{EqFlucInversG} implies that $\delta g^{MN}$ has the following form
\begin{eqnarray}
\delta{g}^{MP}=\delta^{(1)}g^{MP}+\delta^{(2)}{g}^{MP}+\cdots,
\end{eqnarray}
where $\delta^{(n)}g^{MP}$ with a positive integer $n$ is a quantity of order $(H^{MN})^n$. Therefore, Eq.~\eqref{EqFlucInversG} must be satisfied order by order. For the first two orders, we have
\begin{eqnarray}
\label{delg1}
\delta^{(1)} {g}^{MN}&=&-H^{MN},\\
\label{delg2}
\delta^{(2)} g^{MP}&=&H^M_{~N} H^{NP}.
\end{eqnarray}
Similarly, one can derive the first- and second-order perturbations of other geometric quantities, what we need in this paper is the following:
\begin{eqnarray}
\label{pertDet1}
\delta^{(1)}\sqrt{-g}
&=&\frac12\sqrt{-g} H,\\
\label{pertDet2}
\delta^{(2)}\sqrt{-g}
&=&\frac18\sqrt{-g}\left(H^2-2H_{MN}H^{MN}\right),\\
  {\delta ^{(1)}}R &=&{\nabla _M}{\nabla _N}{H^{MN}} - {\nabla _P}{\nabla ^P}H - {H^{MN}}{R_{MN}},\\
\label{pertScaCurv2}
  {\delta ^{(2)}}R &=&  - {H^{MN}}{\nabla ^P}{\nabla _M}{H_{PN}} - {H^{MN}}{\nabla _M}{\nabla ^P}{H_{PN}} + {H^{MN}}{\nabla _P}{\nabla ^P}{H_{MN}} \nonumber \\
   &+& {H^{MN}}{\nabla _M}{\nabla _N}H + H_{A}^M{H^{AN}}{R_{MN}} + {\nabla _M}H{\nabla _N}{H^{MN}} - {\nabla _A}{H^{AN}}{\nabla ^M}{H_{MN}} \nonumber \\
   &+& \frac{3}{4}{\nabla _A}{H_{MN}}{\nabla ^A}{H^{MN}} - \frac{1}{2}{\nabla ^A}{H^{MN}}{\nabla _M}{H_{AN}} - \frac{1}{4}{\nabla _A}H{\nabla ^A}H.
\end{eqnarray}
Note that here we have defined $H\equiv g^{MN}H_{MN}$.

The perturbation of the function $f(\tilde{R})$ can be obtained by simply using the Taylor expansion
\begin{eqnarray}
f(\tilde R) &=& f(R + \delta R) = f(R) + \frac{{df}}{{dR}}\delta R + \frac{1}{2}\frac{{{d^2}f}}{{d{R^2}}}{(\delta R)^2} +  \cdots,
\end{eqnarray}
noting that here
\begin{eqnarray}
\delta R=\delta^{(1)} R+\delta^{(2)} R+\cdots.
\end{eqnarray}
Therefore,
\begin{equation}
\delta f(R)=f(\tilde R)- f(R)= f_R\delta R + \frac{1}{2}f_{RR}{(\delta R)^2} +  \cdots,
\end{equation}
where $f_R=df/dR$ and $f_RR=d^2f/dR^2$.
Obviously,
\begin{eqnarray}
\delta^{(1)} f(R)&=&f_R\delta^{(1)} R,\\
\delta^{(2)} f(R)&=&f_R\delta^{(2)} R + \frac{1}{2}f_{RR}{(\delta^{(1)} R)^2}.
\end{eqnarray}

Then, the first-order variation of the action reads
\begin{eqnarray}
 &&\delta^{(1)} S=2M^{d-2}_\ast\int d^d x [(\delta^{(1)} \sqrt{-{g}})f({R})
 +\sqrt{-{g}}\delta^{(1)} f({R})]\nonumber\\
 &=&2M^{d-2}_\ast\int d^d x \sqrt{-{g}}\bigg( \frac{f(R)}{2}H + {f_R}{\nabla _M}{\nabla _N}{H^{MN}}\nonumber\\
 &-&{f_R}{\nabla _P}{\nabla ^P}H
    - {f_R}{H^{MN}}{R_{MN}} \bigg).
\end{eqnarray}
One can easily prove that the variation of $\delta^{(1)} S$ with respect to $H^{MN}$ simply leads to the background Einstein equations,
\begin{eqnarray}
\label{EqEinstein}
&&{f_R}{R_{MN}} - \frac{1}{2}{g_{MN}}f(R) - {\nabla _M}{\nabla _N}{f_R} + {g_{MN}}{\nabla _P}{\nabla ^P}{f_R}=0.
\end{eqnarray}

Similarly, the master equation for the linear perturbation $H_{\mu\nu}$ can be obtained via the variation of the second-order perturbation of the action, which reads
\begin{eqnarray}
\delta^{(2)}S&=&2M^{d-2}_\ast\int d^d x \{f(R){\delta ^{(2)}}\sqrt { - g}
+ {\delta ^{(1)}}\sqrt { - g} {f_R}{\delta ^{(1)}}R\nonumber\\
&+&
\sqrt { - g} [{f_R}{\delta ^{(2)}}R + \frac{1}{2}{f_{RR}}{({\delta ^{(1)}}R)^2}]\}.
\end{eqnarray}
Using Eqs.~\eqref{pertDet1}-\eqref{pertScaCurv2} and omitting the boundary terms, we finally obtain
\begin{eqnarray}
\label{EqQuadraticCova}
\delta^{(2)}S
&=& M^{d-2}_\ast\int d^d x \sqrt { - g} \bigg\{ \frac{1}{{\rm{4}}}f(R)\left( {{H^2}
   - 2{H_{MN}}{H^{MN}}} \right)+ f_{RR}(\delta^{(1)} R)^2\nonumber\\
&+ &{f_R}\bigg({\nabla _A}{H_{MN}}{\nabla ^M}{H^{AN}}
    - \frac{1}{{\rm{2}}}{\nabla ^A}{H^{MN}}{\nabla _A}{H_{MN}}
    +\frac{1}{{\rm{2}}}{\nabla _A}H{\nabla ^A}H\nonumber\\
&+ &
 {\rm{2}}{H^{MN}}{H^A}_N{R_{MA}} - H{H^{MN}}{R_{MN}} - {\nabla _M}H{\nabla _N}{H^{MN}}\bigg)\nonumber\\
 &+ &{\nabla _A}{f_R}\bigg(2H_{MN}{\nabla ^M}{H^{AN}}
 - {\rm{2}}{H^{AM}}{\nabla ^N}{H_{MN}}\nonumber\\
 &-& H{\nabla _N}{H^{AN}} + H{\nabla ^A}H\bigg)\bigg\}
\end{eqnarray}
Equation~\eqref{EqQuadraticCova} holds for any background geometry. Obviously, the higher-order derivative terms all come from the $f_{RR}(\delta^{(1)} R)^2$ term. Since $R$ is a scalar, the higher-order derivatives appear only in the scalar perturbation equations. If we deal with this term properly, it is possible to obtain second-order perturbation equations. To understand this, we note that the quadratic action Eq.~\eqref{EqQuadraticCova} is invariant under the following gauge transformations (see~\cite{Mukhanov2005} for details):
\begin{eqnarray}
\label{EqGauge1}
\Delta H_{MN}&=&-\nabla_M \xi_N-\nabla_N \xi_M,\\
\label{EqGauge2}
\Delta \delta^{(1)} R&\to&-\xi^M \partial_M R,
\end{eqnarray}
where $\xi^M(x^N)$ is related to the coordinate transformation:
\begin{equation}
x^M \to x^M+ \xi^M(x^N),
\end{equation} and ``$\Delta$" represents the gauge transformation of a linear perturbation.

To fix the gauge freedom, one needs to choose a gauge, or equivalently, to choose a coordinate system $\xi^M$. As we will see below, a suitable choice of gauge can eliminate the $f_{RR}(\delta^{(1)} R)^2$ term, and it leads to simple second-order perturbation equations.

\section{$f(R)$ warped spaces and their linearization}
\label{secWarpedSpace}
As an application of Eq.~\eqref{EqQuadraticCova}, we consider the linearization of an arbitrary five-dimensional warped space, whose metric takes the form~\cite{RandallSundrum1999a}
\begin{eqnarray}
\label{EqMetric}
ds^2=a^2(r)\eta_{MN}dx^M dx^N,
\end{eqnarray}
where $\eta_{MN}=\textrm{diag}(-1,1,1,1,1)$. The unknown function $a(r)$ is called the warp factor, which only depends on the extra dimension $r\equiv x^5$.
In this case, the Einstein equations \eqref{EqEinstein} are
\begin{eqnarray}
\label{EqEinmunu}
  &&a^2
  f(R)+4f_R\left(\frac{a'}{a}\right)^2+2f_R\frac{a''}{a}
  -4f'_R\frac{a'}{a}-2f''_R=0,\\
\label{EqEin55}
 && 8f_R\left(\frac{a'}{a}\right)^2-8f_R\frac{a''}{a}+8f'_R\frac{a'}{a}
  -a^2f(R)=0.
\end{eqnarray}
Here primes denote the derivatives with respect to $r$.

For the background \eqref{EqMetric}, it is convenient to redefine $H_{MN}\equiv a^2h_{MN}$, such that $H^{MN}\equiv g^{MP}g^{NQ}H_{PQ}=a^{-2}h^{MN}$. With these definitions, the indices of $H_{MN}$ (or $h_{MN}$) are raised or lowered with $g_{MN}$ (or $\eta_{MN}$); consequently, $ H\equiv g^{MN}H_{MN}=h$.

One can expand the covariant quadratic action Eq.~\eqref{EqQuadraticCova} in terms of $h_{MN}$ and its partial derivatives. After eliminating all terms that contain $h_{MN}^2$, $h^2$, and $h_{Mr}^2$ by using the background Einstein equations \eqref{EqEinmunu}-\eqref{EqEin55}, the quadratic action finally reduces to
\begin{eqnarray}
\label{EqdeltaS}
&&\delta^{(2)}S= M^{3}_\ast\int d^5 x{a^3}\bigg\{ {f_R}\bigg[{\partial _M}{h_{NP}}{\partial ^P}{h^{MN}} - \frac{1}{2}{\partial _P}{h_{MN}}{\partial ^P}{h^{MN}}
    \nonumber\\&&
   - {\partial ^M}h{\partial ^N}{h_{MN}}
    + \frac{1}{2}{\partial ^P}h{\partial _P}h+ 3\frac{a'}{a}(h{\partial ^\mu }{h_{\mu r}} - h'{h_{rr}})\bigg] \nonumber\\&&
   + f_R'(h{\partial ^\mu }{h_{\mu r}} - h'{h_{rr}}) + f_{RR}(\delta^{(1)} R)^2\bigg\}.
 \end{eqnarray}
Here the indices $\mu,\nu,\cdots$ are raised by the four-dimensional Minkowski metric $\eta^{\mu\nu}$. For convenience we take $M^{3}_\ast=1$ from now on.

\subsection{The Scalar-Tensor-Vector decomposition}
Mathematically, it is always possible to rewrite the metric perturbations into scalar, tensor, and vector modes~\cite{Bardeen1980,KodamaSasaki1984,Weinberg2008,ZhongLiu2013},
\begin{eqnarray}
{h_{\mu r}} &=& {\partial _\mu }F + {G_\mu },\\
{h_{\mu \nu }} &=& {\eta _{\mu \nu }}A + {\partial _\mu }{\partial _\nu }B + 2{\partial _{(\mu }}{C_{\nu )}} + {D_{\mu \nu }},
\end{eqnarray}
%\end{subequations}
where the vector perturbations $C_\mu, G_\mu$ are transverse,
\begin{eqnarray}
\label{ConsVec}
\partial^\mu C_\mu=0=\partial^\mu G_\mu,
\end{eqnarray}
and the tensor perturbation $D_{\mu \nu }$ is transverse and traceless,
\begin{eqnarray}
\label{ConsTens}
\partial^\nu D_{\mu \nu }=0=D^\mu_\mu.
\end{eqnarray}
With these properties, we have
\begin{eqnarray}
  h &=& {h_{rr}} + 4A + {\square ^{(4)}}B, \nonumber \\
  {\partial ^M}{\partial ^N}{h_{MN}} &=& {\partial ^r}{\partial ^r}{h_{rr}} + {\square ^{(4)}}A + {\square ^{(4)}}{\square ^{(4)}}B + 2{\square ^{(4)}}F', \nonumber \\
  {\partial ^M}{h_{Mr}} &=& {\partial ^r}{h_{rr}} + {\square ^{(4)}}F,
\end{eqnarray}
where ${\square ^{(4)}}\equiv \eta^{\mu\nu}\partial_\mu\partial_\nu$.
For convenience, let us define $\psi  = F - \frac{1}{2}B'$ and ${v_\mu } = {G_\mu } - C_\mu '$.

Using the STV decomposition and omitting the total derivative terms, the quadratic action Eq.~\eqref{EqdeltaS} can be rewritten as
\begin{eqnarray}
\label{EqQuadra}
 {\delta ^{(2)}}{S} &=&\frac{1}{2}\int d^5 x{a^3}\bigg\{ {f_R}\bigg[{v^\mu }{\square ^{(4)}}{v_\mu } + \frac{1}{2}{D^{\mu \nu }}{\square ^{(4)}}{D_{\mu \nu }} - \frac{1}{2}{D^{\mu \nu\prime}} D_{\mu \nu }^\prime \nonumber\\&-&
   6A'{\square ^{(4)}}\psi  + 6\frac{a'}{a}{h_{rr}}{\square ^{(4)}}\psi  - 3A{\square ^{(4)}}A - 3{h_{rr}}{\square ^{(4)}}A + 6A'A' \nonumber\\&-&
    3\frac{a'}{a} h_{rr}'{h_{rr}} - 12\frac{a'}{a}A'{h_{rr}}\bigg] + a^2 f_{RR}(\delta^{(1)} R)^2
    \nonumber\\&+&
   2f_R'{h_{rr}}{\square ^{(4)}}\psi  - f_R'h_{rr}'{h_{rr}} - 4f_R'A'{h_{rr}} \bigg\}.
\end{eqnarray}
This is the main result of this paper. Obviously, the vector, tensor and scalar perturbations are decoupled as follows:
\begin{equation}
{\delta ^{(2)}}{S}={\delta ^{(2)}}{S}_{\rm{v}}+
{\delta ^{(2)}}{S}_{\rm{t}}+
{\delta ^{(2)}}{S}_{\rm{s}},
\end{equation}
where the vector and tensor parts are
\begin{eqnarray}
\label{eqvec}
{\delta ^{(2)}}{S}_{\rm{v}}&=&\frac{1}{2}\int d^5 x{\hat{v}^\mu }{\square ^{(4)}}{\hat{v}_\mu},\\
\label{eqtensor}
{\delta ^{(2)}}{S}_{\rm{t}}&=&\frac{1}{4}\int d^5 x{\hat{D}^{\mu\nu} }\left({\square ^{(4)}}{\hat{D}_{\mu\nu}} - \frac{{\lambda ''}}{\lambda }{{\hat D}_{\mu \nu }} + {\hat D}_{\mu \nu }''\right),
\end{eqnarray} respectively.
Here we have defined $\hat{v}_\mu=\lambda  {v}_\mu, ~\hat{D}_{\mu\nu}=\lambda{D}_{\mu\nu}$ and $\lambda= {a^{3/2}}{f_R^{1/2}}$. The same quadratic actions ${\delta ^{(2)}}{S}_{\rm{v}}$ and ${\delta ^{(2)}}{S}_{\rm{t}}$ have also been derived in Ref.~\cite{ZhongLiu2016} by using a conformal transformation.

In order to derive the quadratic action of the scalar perturbations, we need to analyze the gauge degrees of freedom of the total action Eq.~\eqref{EqQuadra}.

\section{The curvature gauge and the scalar normal mode}
\label{secCurv}
With the STV decomposition, the gauge transformation Eq.~\eqref{EqGauge1} can be written as~\cite{ZhongLiu2013}:
\begin{eqnarray}
\Delta A&=&-2\frac {a'}{a}\xi^r, \quad
\Delta h_{rr}=
   -2\xi^{r\prime}
   -2\frac{a'}{a}\xi^r,\nonumber\\
\Delta\psi&=&-\xi^r,\quad \Delta{v_\mu }=0=\Delta D_{\mu\nu},
\end{eqnarray}
and Eq.~\eqref{EqGauge2} reduces to
\begin{equation}
\Delta \delta^{(1)}R=-R'\xi^r,
\end{equation}
where $\xi^r$ is the fifth component of $\xi^M$.

Therefore, the normal modes of the vector and tensor perturbations are already gauge independent. The only gauge degree of freedom lies in the scalar sector. This gauge degree of freedom will be completely fixed after $\xi^r$ is uniquely fixed. The most widely used longitude gauge corresponds to $\xi^r=\psi$, such that in the new coordinate frame $\psi_{\rm{new}}=\psi-\xi^r=0$. In other words, by using the gauge degree of freedom, we can eliminate one of the scalar perturbations. Obviously, in the present work the best choice is to take the curvature gauge such that $\delta^{(1)}R=0$.

Then, the variation of $\delta^{(2)}{S}_{\rm{s}}$ with respect to ${\square^{(4)}}\psi$ would lead to the following constraint:
\begin{eqnarray}
\label{EqCons}
3{f_R}\frac{a'}{a}{h_{rr}} + f_R^\prime {h_{rr}} - 3{f_R}A' = 0.
\end{eqnarray}
Using this relation, we can eliminate $h_{rr}$ in terms of $A$, and
finally get the scalar quadratic action
\begin{equation}
\label{scalarQuard}
  {\delta ^{(2)}}{{S}_{\rm{s}}} = \int d^5x \mathcal{G}\left (\square^{(4)}\mathcal{G}
  +\mathcal{G}''-\frac{\theta''}{\theta}\mathcal{G}\right),
\end{equation}
where
\begin{equation}
\theta=\frac{ a^{3/2} f_R'}{\sqrt{3f_R}\left(\frac{a'}{a}+\frac13\frac{f_R'}{ f_R}\right)},\quad \mathcal{G}=\theta A.
\end{equation}
One can easily check that the $\theta$ defined here is equal to the one defined in Ref.~\cite{ZhongLiu2016} (up to an overall constant coefficient).

From the quadratic actions Eqs.~\eqref{eqvec}-\eqref{eqtensor}, we know that no new degrees come in the tensor and the vector sectors as compared to Einstein's theory. For these two sectors, the only difference between $f(R)$ theory and the Einstein's theory is the factor $\lambda= {a^{3/2}}{f_R^{1/2}}$. In the scalar sector, however, $f(R)$ gravity has one more scalar degree of freedom than in Einstein's theory. Now, let us discuss how this scalar degree comes in.

At the beginning, there are four scalar degrees of freedom: $A, B, E, h_{rr}$. But as we have shown in Eq.~\eqref{EqQuadra}, $B$ and $F$ always appear together: $\psi  = F - \frac{1}{2}B'$.
For Einstein's theory $f_{RR}=0$, $\square^{(4)}\psi$ is simply the Lagrangian multiplier that corresponds to the constraint equation \eqref{EqCons}, and therefore is not a true degree of freedom. We are left with only two scalars $A$ and $h_{rr}$. After using the constraint equation \eqref{EqCons} and fixing the gauge degree of freedom, there is no scalar degree of freedom left. For $f_{RR}\neq 0$, however, $\square ^{(4)}\psi$ is no longer a Lagrangian multiplier because $\delta^{(1)} R$ is a function of $\square ^{(4)}\psi$,
 \begin{eqnarray}
 \label{EqDelR}
  {a^2}\delta^{(1)} R &= & - 16\frac{{a'}}{a}A' - 4A'' + 4\frac{{a'}}{a}{h_{rr}}' + 8\frac{{a''}}{a}{h_{rr}} + 4{\left( {\frac{{a'}}{a}} \right)^2}{h_{rr}} \nonumber \\
   &&+ 8\frac{{a'}}{a}{\square ^{(4)}}\psi  + 2{\square ^{(4)}}\psi'  - {\square ^{(4)}}{h_{rr}} - 3{\square ^{(4)}}A.
\end{eqnarray}
As a consequence, the variation of ${\delta ^{(2)}}{S}$ with respect to $\square ^{(4)}\psi$ leads to the flowing constraint equation:
\begin{eqnarray}
\label{eqCons2}
- 2(f_{RR}\delta^{(1)}R)' + 2\frac{{a'}}{a}f_{RR}\delta^{(1)}R  + 3{f_R}\frac{{a'}}{a}{h_{rr}} + f_R^\prime {h_{rr}} - 3{f_R}A'  = 0,
\end{eqnarray}
which is a relation between $\psi, A$, and $h_{rr}$.
In this case, after using the constraint equation and taking a gauge condition, we are left with one scalar degree of freedom. Therefore, Einstein's theory is a very special case of the $f(R)$ theory where one of the scalar degrees of freedom becomes a Lagrangian multiplier. In general, the equations of scalar perturbations contain third- and fourth-order derivative terms, but under the curvature gauge $\delta^{(1)}R=0$ all these higher-order derivative terms can be completely eliminated. In the end, the master equation for the true scalar perturbation becomes second order.

\section{On the linear equivalence between different frames}
We have shown that the quadratical actions Eqs. \eqref{eqvec}, \eqref{eqtensor} and \eqref{scalarQuard} obtained in the higher-order frame are completely equivalent to those obtained in the Einstein frame (see Ref.~\cite{ZhongLiu2016}). However, it is still unclear to what extent two different frames have equivalent linear structure. For example, if one starts with the widely used longitudinal gauge (see Ref.~\cite{KobayashiKoyamaSoda2002} for a case study of the brane world) in the higher-order and Einstein frames and independently derives the linear perturbation equations, will equivalent results be obtained? This question can also be asked for many other gauges that completely eliminate the gauge degree of freedom. In the present paper, we only offer an explanation to the question: Why dose the curvature gauge in the higher-order frame lead to the same quadratic actions as those obtained in the Einstein frame~\cite{ZhongLiu2016}?

Since the vector and the tensor perturbations are already gauge independent, and the conformal transformation only redefines these perturbations, it should not be strange that the final equations obtained in different frames coincide with each other. For this reason, we only discuss the scalar sector.

Usually, there are many gauge choices which can eliminate the gauge degrees of freedom, and for each of these gauges one can define a group of gauge-invariant variables.  Thus, in order to eliminate the gauge degrees of freedom, one can either choose a gauge or use the corresponding gauge-invariant variables. These two approaches lead to equivalent results. However, from the point of view of quadratic actions, only one group of the gauge-invariant variables can finally diagonalize the scalar quadratic action~\cite{Giovannini2001a,ZhongLiu2013}. In the case of the vacuum $f(R)-$brane (or the minimally coupled scalar-Einstein theory), only one of the gauge-invariant variables is independent. This gauge-invariant variable corresponds to the normal mode of the scalar sector, and its quadratic action will be the starting point when one considers the quantization of the perturbation. By definition, the gauge invariant quantity $\hat{\mathcal{G}}$ defined in Ref.~\cite{ZhongLiu2016} is the normal mode of the scalar perturbation for the $f(R)-$brane in the Einstein frame.

For the curvature gauge used in the present paper, we can also define a gauge-invariant quantity,
\begin{equation}
\mathcal{R}\equiv A- 2\frac{a'}{aR'}\delta R.
\end{equation}
The quadratic action Eq.~\eqref{scalarQuard} can be expressed in terms of $A$ by taking the gauge condition $\delta R=0$, as we have done in this paper, or equivalently, in terms of $\mathcal{R}$ without taking the curvature gauge. Both approaches are mathematically equivalent since under the gauge $\delta R=0$ we simply have $\mathcal{R}= A$. In other words, $\mathcal{R}$ (or $A$) is the scalar normal mode of the $f(R)-$brane in the higher-order frame\footnote{Equivalently, one can also take the gauge $A=0$ and regard $\delta R$ as the normal mode.}. The study of our work only shows that the quadratic actions of the normal modes (including scalar, tensor and vector modes) are equivalent for these two frames. It will be interesting to consider the frame dependence for other gauge choices, especially the longitudinal gauge which has been widely used in literature, but we leave this issue for our future works.

\section{Summary}
\label{secSum}
In this work, we considered the linearization of a five-dimensional warped $f(R)$ gravity in the higher-order frame. We first derived the quadratic action of the metric perturbation around any background. Then we focused on the case of a five-dimensional warped space. We showed that by using the STV decomposition, the quadratic action can be separated into scalar, vector, and  tensor sectors. The vector and tensor parts are already gauge independent, and gauge choice is needed only in the scalar sector. Instead of the longitude gauge, we applied the curvature gauge, which eliminates all the higher-order terms in the quadratic actions, and the residual scalar mode $A$ can be regarded as the scalar normal mode. As compared to the quadratic actions for the normal modes in the Einstein frame~\cite{ZhongLiu2016}, we found that the quadratic actions for the normal modes (including scalar, tensor and vector sectors) are equivalent in these two frames. Since in literature the longitudinal gauge was used widely, it is also important to discuss whether the master equation corresponding to the longitudinal gauge as well as many other gauges is frame independent or not. However, this question is beyond the scope of the present work, and we leave it for future work.

This work also sets an simple example on how to linearize a higher-order gravity without using conformal transformation, and it would be useful when more complicated higher-order gravitational theories are considered.
\section*{Acknowledgments}
This work was supported by the National Natural Science
Foundation of China (Grants No. 11375075, No. 11522541
and No. 11605127). Yuan Zhong was also supported by China
Postdoctoral Science Foundation (Grant No. 2016M592770).

\section*{References}
%merlin.mbs 2010-03-15 4.21a (PWD, AO, DPC)
%Control: key (0)
%Control: author (8) initials jnrlst
%Control: editor formatted (1) identically to author
%Control: production of article title (-1) disabled
%Control: page (0) single
%Control: year (1) truncated
%Control: production of eprint (0) enabled
%

%\bibliographystyle{apsrev4-1}
%\bibliography{D:/jabref/library/articles/bib/articles}
\end{document}